\documentclass[iop]{emulateapj}
\usepackage{gensymb} \usepackage{subfigure}
\DeclareRobustCommand{\orderof}{\ensuremath{\mathcal{O}}}
\RequirePackage{lineno}
\usepackage{color,soul}
\usepackage{amsmath}
\sethlcolor{yellow}
\begin{document}
\title{Search for a correlation between very-high-energy gamma rays
  and giant radio pulses in the Crab pulsar}

\author{
E.~Aliu\altaffilmark{1},
S.~Archambault\altaffilmark{2},
T.~Arlen\altaffilmark{3},
T.~Aune\altaffilmark{4},
M.~Beilicke\altaffilmark{5},
W.~Benbow\altaffilmark{6},
A.~Bouvier\altaffilmark{4},
J.~H.~Buckley\altaffilmark{5},
V.~Bugaev\altaffilmark{5},
K.~Byrum\altaffilmark{7},
A.~Cesarini\altaffilmark{8},
L.~Ciupik\altaffilmark{9},
E.~Collins-Hughes\altaffilmark{10},
M.~P.~Connolly\altaffilmark{8},
W.~Cui\altaffilmark{11},
R.~Dickherber\altaffilmark{5},
C.~Duke\altaffilmark{12},
J.~Dumm\altaffilmark{13},
A.~Falcone\altaffilmark{14},
S.~Federici\altaffilmark{15,16},
Q.~Feng\altaffilmark{11},
J.~P.~Finley\altaffilmark{11},
G.~Finnegan\altaffilmark{17},
L.~Fortson\altaffilmark{13},
A.~Furniss\altaffilmark{4},
N.~Galante\altaffilmark{6},
D.~Gall\altaffilmark{18},
G.~H.~Gillanders\altaffilmark{8},
S.~Godambe\altaffilmark{17},
S.~Griffin\altaffilmark{2},
J.~Grube\altaffilmark{9},
G.~Gyuk\altaffilmark{9},
D.~Hanna\altaffilmark{2},
J.~Holder\altaffilmark{19},
H.~Huan\altaffilmark{20},
G.~Hughes\altaffilmark{15},
T.~B.~Humensky\altaffilmark{21},
P.~Kaaret\altaffilmark{18},
N.~Karlsson\altaffilmark{13},
Y.~Khassen\altaffilmark{10},
D.~Kieda\altaffilmark{17},
H.~Krawczynski\altaffilmark{5},
F.~Krennrich\altaffilmark{22},
M.~J.~Lang\altaffilmark{8},
S.~LeBohec\altaffilmark{17},
K.~Lee\altaffilmark{5},
M.~Lyutikov\altaffilmark{11},
A.~S~Madhavan\altaffilmark{22},
G.~Maier\altaffilmark{15},
P.~Majumdar\altaffilmark{3},
S.~McArthur\altaffilmark{5},
A.~McCann\altaffilmark{23,32},
P.~Moriarty\altaffilmark{24},
R.~Mukherjee\altaffilmark{1},
T.~Nelson\altaffilmark{13},
A.~O'Faol\'{a}in de Bhr\'{o}ithe\altaffilmark{10},
R.~A.~Ong\altaffilmark{3},
M.~Orr\altaffilmark{22},
A.~N.~Otte\altaffilmark{25,32},
N.~Park\altaffilmark{20},
J.~S.~Perkins\altaffilmark{26,27},
M.~Pohl\altaffilmark{16,15},
H.~Prokoph\altaffilmark{15},
J.~Quinn\altaffilmark{10},
K.~Ragan\altaffilmark{2},
L.~C.~Reyes\altaffilmark{28},
P.~T.~Reynolds\altaffilmark{29},
E.~Roache\altaffilmark{6},
D.~B.~Saxon\altaffilmark{19},
M.~Schroedter\altaffilmark{6,32},
G.~H.~Sembroski\altaffilmark{11},
G.~D.~\c{S}ent\"{u}rk\altaffilmark{21},
A.~W.~Smith\altaffilmark{17},
D.~Staszak\altaffilmark{2},
I.~Telezhinsky\altaffilmark{16,15},
G.~Te\v{s}i\'{c}\altaffilmark{2},
M.~Theiling\altaffilmark{11},
S.~Thibadeau\altaffilmark{5},
K.~Tsurusaki\altaffilmark{18},
A.~Varlotta\altaffilmark{11},
S.~Vincent\altaffilmark{15},
M.~Vivier\altaffilmark{19},
R.~G.~Wagner\altaffilmark{7},
S.~P.~Wakely\altaffilmark{20},
T.~C.~Weekes\altaffilmark{6},
A.~Weinstein\altaffilmark{22},
R.~Welsing\altaffilmark{15},
D.~A.~Williams\altaffilmark{4},
B.~Zitzer\altaffilmark{7},
V.~Kondratiev\altaffilmark{30,31}
}

\altaffiltext{1}{Department of Physics and Astronomy, Barnard College, Columbia University, NY 10027, USA}
\altaffiltext{2}{Physics Department, McGill University, Montreal, QC H3A 2T8, Canada}
\altaffiltext{3}{Department of Physics and Astronomy, University of California, Los Angeles, CA 90095, USA}
\altaffiltext{4}{Santa Cruz Institute for Particle Physics and Department of Physics, University of California, Santa Cruz, CA 95064, USA}
\altaffiltext{5}{Department of Physics, Washington University, St. Louis, MO 63130, USA}
\altaffiltext{6}{Fred Lawrence Whipple Observatory, Harvard-Smithsonian Center for Astrophysics, Amado, AZ 85645, USA}
\altaffiltext{7}{Argonne National Laboratory, 9700 S. Cass Avenue, Argonne, IL 60439, USA}
\altaffiltext{8}{School of Physics, National University of Ireland Galway, University Road, Galway, Ireland}
\altaffiltext{9}{Astronomy Department, Adler Planetarium and Astronomy Museum, Chicago, IL 60605, USA}
\altaffiltext{10}{School of Physics, University College Dublin, Belfield, Dublin 4, Ireland}
\altaffiltext{11}{Department of Physics, Purdue University, West Lafayette, IN 47907, USA }
\altaffiltext{12}{Department of Physics, Grinnell College, Grinnell, IA 50112-1690, USA}
\altaffiltext{13}{School of Physics and Astronomy, University of Minnesota, Minneapolis, MN 55455, USA}
\altaffiltext{14}{Department of Astronomy and Astrophysics, 525 Davey Lab, Pennsylvania State University, University Park, PA 16802, USA}
\altaffiltext{15}{DESY, Platanenallee 6, 15738 Zeuthen, Germany}
\altaffiltext{16}{Institute of Physics and Astronomy, University of Potsdam, 14476 Potsdam-Golm, Germany}
\altaffiltext{17}{Department of Physics and Astronomy, University of Utah, Salt Lake City, UT 84112, USA}
\altaffiltext{18}{Department of Physics and Astronomy, University of Iowa, Van Allen Hall, Iowa City, IA 52242, USA}
\altaffiltext{19}{Department of Physics and Astronomy and the Bartol Research Institute, University of Delaware, Newark, DE 19716, USA}
\altaffiltext{20}{Enrico Fermi Institute, University of Chicago, Chicago, IL 60637, USA}
\altaffiltext{21}{Physics Department, Columbia University, New York, NY 10027, USA}
\altaffiltext{22}{Department of Physics and Astronomy, Iowa State University, Ames, IA 50011, USA}
\altaffiltext{23}{Kavli Institute for Cosmological Physics, University of Chicago, Chicago, IL 60637, USA}
\altaffiltext{24}{Department of Life and Physical Sciences, Galway-Mayo Institute of Technology, Dublin Road, Galway, Ireland}
\altaffiltext{25}{School of Physics and Center for Relativistic Astrophysics, Georgia Institute of Technology, 837 State Street NW, Atlanta, GA 30332-0430}
\altaffiltext{26}{CRESST and Astroparticle Physics Laboratory NASA/GSFC, Greenbelt, MD 20771, USA.}
\altaffiltext{27}{University of Maryland, Baltimore County, 1000 Hilltop Circle, Baltimore, MD 21250, USA.}
\altaffiltext{28}{Physics Department, California Polytechnic State University, San Luis Obispo, CA 94307, USA}
\altaffiltext{29}{Department of Applied Physics and Instrumentation, Cork Institute of Technology, Bishopstown, Cork, Ireland}
\altaffiltext{30}{ASTRON, The Netherlands Institute for Radio Astronomy, Postbus 2, 7990 AA, Dwingeloo, The Netherlands}
\altaffiltext{31}{Astro Space Center of the Lebedev Physical Institute, Profsoyuznaya str. 84/32, Moscow 117997, Russia}
\altaffiltext{32}{To whom correspondence should be addressed. E-mail: mccann@kicp.uchicago.edu (A.M.); nepomuk.otte@gmail.com (A.N.O.); schroedter@veritas.sao.arizona.edu (M.S.)}

\begin{abstract}
We present the results of a joint observational campaign between the
Green Bank radio telescope and the VERITAS gamma-ray telescope, which
searched for a correlation between the emission of very-high-energy
(VHE) gamma rays ($E_{\gamma} >$ 150~GeV) and Giant Radio Pulses
(GRPs) from the Crab pulsar at 8.9~GHz. A total of 15366 GRPs were
recorded during 11.6 hours of simultaneous observations, which were
made across four nights in December 2008 and in November and December
2009. We searched for an enhancement of the pulsed gamma-ray emission
within time windows placed around the arrival time of the GRP
events. In total, 8 different time windows with durations ranging from
0.033~ms to 72~s were positioned at three different locations relative
to the GRP to search for enhanced gamma-ray emission which lagged,
led, or was concurrent with, the GRP event. Further, we performed
separate searches on main pulse GRPs and interpulse GRPs and on the
most energetic GRPs in our data sample.  No significant enhancement of
pulsed VHE emission was found in any of the preformed searches. We set
upper limits of 5-10 times the average VHE flux of the Crab pulsar on
the flux simultaneous with interpulse GRPs on single-rotation-period
time scales. On $\sim$8-second time scales around interpulse GRPs, we
set an upper limit of 2-3 times the average VHE flux. Within the
framework of recent models for pulsed VHE emission from the Crab
pulsar, the expected VHE-GRP emission correlations are below the
derived limits.
\end{abstract}


\section{Introduction}
The Crab pulsar, PSR B0531+21, is a powerful young pulsar and one of
the most studied objects in the sky. It is one of the brightest
pulsars in the high-energy gamma-ray regime \citep{Fierro98,Abdo2010}
and the only pulsar so far to be detected above 100 GeV
\citep{Aliu2011Sci,MAGIC2011}. The Crab pulsar is also one of only
several known pulsars which exhibit the giant radio pulse phenomenon
\citep{Knight06}: single radio pulses with flux densities that greatly
exceed the average pulse flux density and can, at their maximum, be as
bright as a few million Janskys \citep{Soglasnov07}. The energies of
giant pulses follow a power-law distribution
\citep{Cordes2004,Popov2007} in contrast to the regular pulses that
obey a Gaussian or log-normal distribution
\citep{BurkeSpolaor2012}. High-resolution time measurements of
individual GRPs reveal pulses which can be a few microseconds to a few
nanoseconds wide \citep{Hankins2003Natur}, with the narrower GRPs
possessing the highest flux density.

The pulse profile of the Crab pulsar is dominated across the
electromagnetic spectrum, from radio waves to VHE gamma rays, by two
emission peaks, referred to as the main pulse (MP) and interpulse
(IP). These peaks occur at phases 0.0 and 0.4, respectively, with a
``bridge'' of enhanced emission appearing between these two peaks in
the optical, X-ray and gamma-ray pulse profiles
\citep{Oosterbroek08,Mineo06,Kuiper01,Abdo2010}. Above 100 GeV,
significant emission is only observed during the main and inter-pulse
\citep{Aliu2011Sci,MAGIC2011}. In the radio-pulse profile several
different components appear at different frequencies, including a
precursor to the main radio pulse and two ``high frequency
components'' (HFCs), which only appear above $\sim$5 GHz and occur at
phases $\sim$0.77 and $\sim$0.93 \citep{Moffett96}. In the Crab
pulsar, giant radio pulses have been observed to occur during both the
main pulse and the interpulse. Indeed, it has been suggested that the
Crab radio profile is composed of entirely GRPs, with the ``regular''
pulsar pulse corresponding to the main pulse precursor component
\citep{Popov2006}. The case for the regular pulse is difficult to
disentangle, since single regular pulses cannot be observed over the
nebular background. Stark differences in the characteristics of main
pulse GRPs and interpulse GRPs have been revealed by observations of
the Crab pulsar between 6 and 10.5 GHz. In this frequency range,
interpulse GRPs are typically several microseconds long and populate a
set of regularly spaced frequency bands \citep{Hankins2007}. On the
other hand, main pulse GRPs exhibit broadband spectra and appear as a
succession of narrow pulses ranging from unresolved widths
  below 0.4~ns to widths of a few microseconds
\citep{Hankins2003Natur,Hankins2007}. These striking differences
suggest that the emission mechanisms may differ between the
main pulse GRPs and interpulse GRPs in the Crab pulsar above 5 GHz.

The mechanisms responsible for the generation of GRPs are still
unknown. Changes in the coherence of the plasma beam, which is
believed to be responsible for the normal pulsed radio emission, can
in principle explain the generation of GRPs. However, mechanisms which
primarily affect the coherence of the plasma have no effect on the
incoherent emission from pulsars and, thus, no enhancement is expected
in the gamma-ray emission in connection with GRP events. Mechanisms
which increase the rate of particle production within the
magnetospheric emission region, or which change the direction of the
emission beam, should, however, affect the higher energy incoherent
emission. Such mechanisms may create an enhancement in the gamma-ray
emission from the pulsar.

The exceptional emission band structure seen in interpulse GRPs from
the Crab pulsar at high frequencies reported by \cite{Hankins2007}
prompted a quantitative model for their generation by
\cite{Lyutikov2007}. This model argues that GRPs are generated in a
dense plasma region close to the last closed magnetic field
line. Occasional magnetic field reconnections excite the plasma
producing a particle beam with a high Lorentz factor which emits a GRP
via anomalous cyclotron resonance. This model, when fit to the
interpulse GRP data from \cite{Hankins2007}, returns a predicted
Lorentz factor for the beam, $\gamma = \orderof(10^{8})$, which is
large enough to cause the particles within the beam to generate
curvature photons with energies as high as tens of GeV. Thus, a
feature of this model is a prediction of an enhancement in the
gamma-ray emission correlated with high frequency interpulse GRPs from
the Crab pulsar.

Several studies have been performed to examine a possible connection
between GRPs in the Crab and higher-energy incoherent
emission. \cite{Lundgren1995} studied 50-220 keV gamma rays recorded
by the OSSE instrument on board NASA's \emph{Compton Gamma-ray
  Observatory} along side Green Bank radio data recorded at 800, 812.5
and 1330 MHz. No enhancement was seen in the gamma-ray emission on
2-minute time scales concurrent with GRP events. This study yielded an
upper limit on the average gamma-ray flux from the pulsar over 2
minutes concurrent with GRPs of 2.5 times the average pulsed gamma-ray
flux. Based on the non-detection of a gamma-ray enhancement, the
authors argue in favour of coherence changes within the emission
plasma as the source of GRP generation. \cite{Shearer2003}, however,
observed a significant 3\% enhancement in the optical main pulse
concurrent with the period of emission of GRPs measured at 1380
MHz. No enhancement was seen in the interpulse. This observation
suggests a link between coherent radio emission and incoherent optical
emission in the Crab pulsar. Small changes in the pair-creation rate
leading to localised density increases within the emission plasma
could create the GRP event and provide a small enhancement in the
optical incoherent emission.

Recently \cite{Bilous2011} studied 0.1-5 GeV gamma rays recorded by
the \emph{Fermi}-LAT in conjunction with 8.9 GHz observations of the
Crab performed by the Green Bank radio telescope. No enhancement was
seen in the gamma-ray flux within the main pulse, interpulse and
bridge, during single rotation periods which contained a GRP. This
yielded a limit on the flux from the pulsar during rotation periods
concurrent with GRPs of less than 4 times the average gamma-ray flux
from the pulsar. This result suggests that enhanced pair-creation is
not the dominant factor in the emission of GRPs in the Crab and
supports the idea that local coherence changes in the magnetospheric
plasma cause GRPs, or, that the putative enhanced gamma-ray emission
is beamed in a different direction than the radio emission and thus
unobservable from Earth. Another recent study by \cite{Bilous2012},
which examined 1.4-4.5 keV X-rays from \emph{Chandra} observations of
the Crab pulsar in conjunction with radio observations by the Green
Bank telescope at 1.1-1.9 GHz, found no enhancement in the X-ray flux
in connection with GRPs. This study yielded limits on the flux
enhancement in the main pulse and interpulse concurrent with GRP
events of 10\% and 30\%, respectively.

Along with the aforementioned studies which searched for a correlation
between GRPs and incoherent emission below 10~GeV, a GRP correlation
search was performed on the Crab pulsar at TeV energies in the early
1970s \citep{Argyle74}. This pioneering attempt was performed with the
25-m DRAO radio telescope and the Whipple gamma-ray telescope almost
two decades before the first conclusive detection of a TeV gamma-ray
source \citep{Weekes89}. No TeV-GRP correlation was observed and we
are aware of no other study on this subject with subsequent TeV
instruments.

VERITAS is an array of four imaging atmospheric Cherenkov gamma-ray
telescopes located at the Fred Lawrence Whipple Observatory at the
base of Mount Hopkins in southern Arizona \citep{Holder2006}.  Each
telescope uses a 12~m Davies-Cotton reflector \citep{DaviesCotton}
instrumented with a photo-multiplier tube (PMT) camera with 499
pixels.  The array is sensitive to gamma rays with energies between
100 GeV and 30 TeV. The recent detection by VERITAS of pulsed emission
above 100 GeV from the Crab \citep{Aliu2011Sci} is not expected within
the context of contemporary pulsar models in which curvature radiation
is the dominant gamma-ray emission mechanism. The origin of the
power-law extension of the spectral energy distribution above 100 GeV
seen now by both VERITAS and MAGIC \citep{MAGIC2011} is poorly
understood. A possible explanation is that the VHE emission is caused
by inverse-Compton (IC) up-scattering of soft photons by particles
accelerated in the outer magnetosphere
\citep{Romani96,Lyutikov2011,Du2012}.  Another possible explanation is
that the pulsed VHE emission is produced when pulsed magnetospheric
X-ray photons are up-scattered by relativistic plasma in the wind
outside the light cylinder \citep{Aharonian2012}. Given that the
origin of the VHE emission and the generation of GRPs in the Crab
pulsar are both poorly understood energetic phenomena, we are
motivated to probe whether they are connected.

The remainder of this paper is structured in the following way. In
$\S2$ we describe the simultaneous observations of the Crab pulsar
made with the GBT and VERITAS gamma-ray telescopes. We also describe
the extraction of GRP events from the radio dataset. In $\S3$ we
discuss the gamma-ray data processing and our strategy for searching
for correlated emission within the radio and gamma-ray datasets. In
$\S4$ we describe a Monte-Carlo simulation of the Crab pulsar signal
within the VERITAS dataset, which is used to calculate the strength of
any correlation found in the datasets. In $\S5$ we detail the results
of the enhancement search and in $\S6$ we provide some concluding
remarks.

\section{Observations}
\begin{figure}
\centering \includegraphics[width=0.45\textwidth]{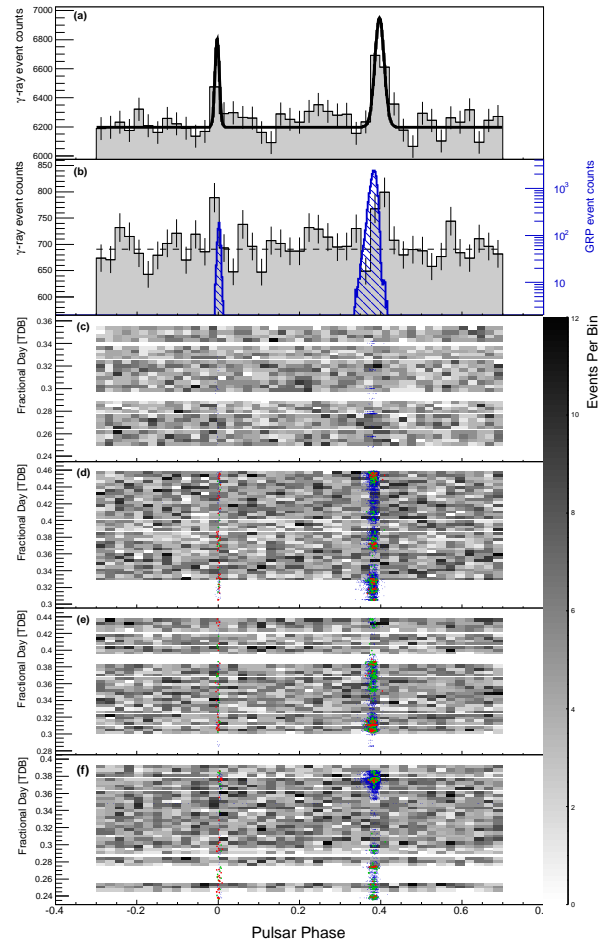}
\caption{Summary of the gamma-ray and radio observations. Panel (a)
  shows the gamma-ray pulse profile for the published 107 hour VERITAS
  Crab Pulsar dataset \citep{Aliu2011Sci}. The overlaid solid line is
  the best fit function determined from a maximum likelihood fit to
  the unbinned VERITAS data. Panel (b) shows the gamma-ray (grey) and
  giant radio pulse (hatching) profiles for the four nights of
  simultaneous observations. Panels (c),(d),(e) and (f) show the
  arrival time at the barycenter versus phase for the observation
  dates 54829, 55153, 55158 and 55180 in MJD respectively. The
  grey-scale histograms show the VERITAS data while the blue, green
  and red points are the radio data with energies less than
  100~Jy$\mu$s, greater than 100~Jy$\mu$s and greater than
  150~Jy$\mu$s, respectively (see colour figure in on-line version).}
\label{fig:summary}
\end{figure}

\begin{table*}
\begin{center}
\begin{tabular}{ccccccc}
Date       & \# GRPs & \# GRPs   & \# $\gamma$-ray & T Overlap  & Center Freq. & $\Delta$ samp. \\ 
$[$MJD]    &         & Overlap   & Candidates      & [min]      & [MHz]        & [$\mu$s]\\\hline 
54829      & 280     & 150       &  5566           & 138.365    & 8832         & 40.96\\
55153      & 7813    & 5937      &  8738           & 200.232    & 8900         &  3.2\\
55158      & 3771    & 3243      &  7928           & 182.881    & 8900         &  3.2\\
55180      & 6916    & 6036      &  7861           & 180.107    & 8900         &  3.2\\\hline
Total      & 18780   & 15366     & 30093           & 701.585    &              &
\end{tabular}
\end{center}
\caption{A summary of the radio and gamma-ray data used in this
  study. Measurements were made on four separate nights with a total
  11.6 hours of simultaneous data accumulated. The significantly
  smaller number of GRP events found on the first night of
  observations is likely due to higher attenuation of the radio input
  signal. }
\label{tab:GPsummary}
\end{table*}
Simultaneous observations of the Crab pulsar were made by VERITAS and
the 100-m Robert C. Byrd Green Bank Telescope (GBT) on the 29th of
December 2008, the 18th and 23rd of November 2009 and the 15th of
December 2009.  A total of 11.6 hours of simultaneous data was
recorded across these four nights with 2.3, 3.33, 3.04 and 3.00 hours
acquired, respectively. Table~\ref{tab:GPsummary} gives a summary of
the data used in this study.

\subsection{GBT Observations and GRP selection}
The radio data presented here were acquired using the Green Bank
Ultimate Pulsar Processor Instrument (GUPPI) in search mode. The total
bandwidth of 800~MHz was centered at 8832 or 8900~MHz and split into
128 or 256 frequency channels for our observations in 2008 and 2009,
respectively. Full Stokes parameters were recorded at a sampling
interval of $40.96~\mu$s in 2008, whereas total intensity was recorded
with a sampling interval of $3.2~\mu$s in our 2009 observations.

Recorded data from every session were dedispersed using the PRESTO
pulsar software
package\footnote{\url{http://www.cv.nrao.edu/$\sim$sransom/presto/}},
and searched for all single-pulse events with a peak signal-to-noise
ratio seven times greater than the average radio signal. The
dedispersion was done using the contemporaneous dispersion measure
(DM) value stated in the Jodrell Bank Crab pulsar monthly
ephemeris\footnote{\url{http://www.jb.man.ac.uk/pulsar/crab.html}}
\citep{Lyne1993}. The GRP selection was performed with the PRESTO tool
\texttt{singlepulse\_search$.$py}, which convolves the dedispersed
time series with a series of boxcar functions of different widths. The
times of arrival of the selected GRP events were converted into
Tempo\footnote{\url{http://tempo.sourceforge.net}} format and
transformed to barycentric dynamical time (TDB) for the correlation
analysis with the VERITAS events.
 
Between the observation sessions in 2008 and 2009, the DM has
increased from $56.7883\,$~pc\,cm$^{-3}$ in December 15, 2008 to
$56.8279\,$~pc\,cm$^{-3}$ in December 2009. We estimate timing errors
at our high observing frequency of 8.9~GHz to be $\sim0.4~\mu$s, which
is less than our sampling time. Here we assume a smooth variation in
the DM and that the changes in the DM between our observing sessions
are certainly less than the measured change of the DM of $\approx
0.04\,$~pc\,cm$^{-3}$ over the course of one year.

The system equivalent flux density is mostly determined by the Crab
Nebula.  Flux densities of the Crab Nebula were calculated with the
relation $S(f) = 955\times(f/\mathrm{GHz})^{-0.27}$~Jy
\citep{Cordes2004}, accounting for the fact that at 8.9 GHz the solid
angle of the GBT beam covers only 6.25\% of the area occupied by the
nebula. We estimate the system equivalent flux density to be about
0.2~Jy for our 2008 observations, and about 0.7~Jy for the observing
session in 2009. The smaller number of GRPs found in the 2008 dataset
is likely due to a higher attenuation in the receiver system in 2008
where two back-end readouts were used together; the GBT Pulsar Spigot
Card and the GBT Mk5 disk readout. The clumping of the barycentric
arrival times of the GRPs clearly seen in the 2009 observations (see
Figures~1 and 4) agrees with what one would expect from refractive
interstellar scintillations (RISS). At 8.9~GHz the characteristic time
scale of RISS is about 80~minutes \citep{Bilous2011}.

\subsection{VERITAS Observations}
The VERITAS observations were made under the best possible sky
conditions with each of the four telescopes fully operational. These
observations were made with zenith angles smaller than $30^{\degree}$
and an average zenith angle of $16^{\degree}$, yielding the lowest
possible energy threshold. All data were acquired in \textit{wobble}
mode \citep{Fomin1994}. This mode of observation places the object
under study at an offset of $0.5^{\degree}$ from the centre of the
field of view of the telescope, allowing for a simultaneous
measurement of the background in other regions of the field of view
with the same acceptance as the region containing the source.

Before an event is recorded with VERITAS, at least two of the four
telescopes must trigger on a Cherenkov flash within 50~ns. A single
telescope trigger is formed when three or more adjacent PMTs in the
camera register at least six photoelectrons within 9~ns. When the
trigger condition is satisfied, the PMT traces in each telescope are
read out by 500 mega-sample per second flash analogue-to-digital
converters which are located at each telescope. For each
event, the VERITAS data stream contains the GPS time stamp from each
of the four telescopes along with the digitised PMT-traces. In the
entire VERITAS dataset presented in this work, the four GPS timestamps
never diverged by more than 10~$\mu$s, providing a sound basis for
correlation and timing analyses.

\section{Data Analysis}
\subsection{Gamma-ray Event Selection}
Prior to performing a correlation search between the gamma-ray and
radio datasets, the VERITAS data were passed through an analysis
pipeline, which reconstructs the arrival direction and the energy of
the gamma-ray candidate from the Cherenkov images recorded by the
telescopes. For each event, the signal in each PMT is corrected for
gain differences between the PMTs and signals which only contain noise
are removed. Following this image cleaning, images which have a
combined PMT signal (\textit{size}) above 20 photoelectrons have their
second moments calculated about their major and minor axes
\citep{Hillas85}. The arrival direction and impact parameter of the
gamma-ray candidates are then calculated from the intersection points
of the major axes of the images from multiple telescopes when
projected on the sky and ground planes respectively
\citep{Hofmann99}. Background suppression, including cosmic-ray
rejection, is performed by comparing measured event parameters with
Monte-Carlo gamma-ray simulations. Selection parameters include image
brightness and shape, arrival direction, height of shower maximum, and
impact distance, and are combined in multi-dimensional
energy-dependent look-up tables \citep{Krawczynski06}. The optimal cut
values were chosen a priori by modelling the gamma-ray signal from the
Crab pulsar as a power-law with an integral flux of 1\% of the Crab
Nebula flux above 100 GeV and a spectral index of $-4$. A grid search
for the optimal gamma-ray cut values was then performed using this
simulated signal with real Crab Nebula data as background. The
gamma-ray selection parameters and the resulting optimal cut values
are: the angular separation between the source location and the shower
direction, \textit{theta} ($<0.27^{\degree}$), \textit{mean scaled
  width} ($<$1.17), \textit{mean scaled length} ($<$1.35), and height
of shower maximum ($>$6.6~km). This cut optimisation yielded an energy
threshold of 120~GeV for sources with a spectral index of -4.  This
gamma-ray analysis is identical to the analysis which was used in our
paper on the first detection of the Crab pulsar above 100~GeV
\citep{Aliu2011Sci}. The GPS time of the candidate gamma-ray events
which passed cuts was converted to barycentric dynamical time and
phase-folded using the Crab pulsar monthly timing ephemeris with an
in-house software package. The accuracy of these calculations was
cross-checked using the Tempo program.

\subsection{Correlation Search Strategy}
\begin{figure}
\centering \includegraphics[width=0.45\textwidth]{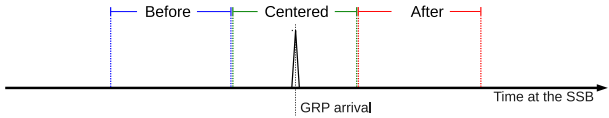}
\caption{Positioning of the gamma-ray search windows with respect to
  the time of arrival of the GRP event at the solar system barycentre
  (SSB). For each search window duration, a window was positioned
  centered on the GRP, with leading search and lagging search windows
  placed directly before and directly after the centered search
  window, respectively.}
\label{fig:windows}
\end{figure}
The physical mechanisms which are responsible for the emission of GRPs
are not well understood. The nature of a possible connection between
GRPs and the formation of high-energy incoherent emission from pulsars
is also unknown. Given these unknowns, we prepared a correlation
search strategy which probed for a connection between GRPs and
very-high-energy gamma-ray emission on different time scales, allowing
for lagging or leading between gamma-ray and GRP emission. The
searches were done by selecting only those gamma-ray events which
arrived at the solar system barycenter within a certain time window
around a GRP event. The durations of the time windows were specified
in units of pulsar rotations. We employed eight different time window
durations lasting; 1, 3, 9, 27, 81, 243, 729 and 2187 pulsar
rotations. This covers time intervals from 0.033 to 72.17 seconds in
time spacings which scale as log to the base 3. Each time window was
used to search for enhanced gamma-ray emission which lagged, led or
was contemporaneous with, the GRP. Thus, for each search window
duration, a window was positioned centered on the GRP, with the
leading and lagging windows placed directly before and directly after
the centered window, respectively (see
Figure~\ref{fig:windows}). Finally, searches were performed
considering only main~pulse GRPs, only interpulse GRPs and both main
and interpulse GRPs combined. Thus, a total of 72 searches were
performed.

In each of the 72 searches, the phase of the gamma-ray events which
fall within the search window is calculated. In our earlier
measurement of pulsed VHE emission from the Crab pulsar
\citep{Aliu2011Sci}, we determined the phases of emission to be
between $-0.013$ and 0.009 for the main pulse and between 0.375 and
0.421 for the interpulse. In this study we only consider VERITAS
events that fall within these phase regions. When a gamma-ray event
falls within the search window defined by a main pulse GRP, it has to
lie within the VHE main pulse emission phase range for it to be
considered in the enhancement search. Corresponding selection criteria
are applied to those gamma-ray events which fall within a window
defined by an interpulse GRP. In the searches which consider both main
and interpulse GRPs, gamma-ray events which fall within either of the
VHE emission phase ranges are selected. The prescription for these 72
searches was defined before the datasets were analysed.

\section{Monte-Carlo Time Series}

\begin{figure}
\centering \includegraphics[width=0.45\textwidth]{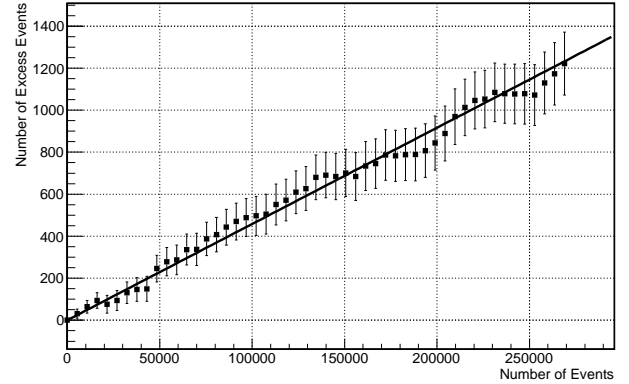}
\caption{The number of excess gamma-ray events in the phase interval
  $-0.013$ to 0.009 (main pulse) and 0.375 to 0.421 (interpulse)
  plotted against the total number of events for the complete 107 hour
  VERITAS dataset \citep{Aliu2011Sci}. The excess growth is clearly
  linear. From a total excess of 1256$\pm$130 in 267088 events, there
  are 4.7$\pm$0.48 excess events in the VHE emission phases per 1000
  events selected.}
\label{fig:excess}
\end{figure}

In order to determine the presence of an enhancement in the VHE
emission from the pulsar correlated with GRPs, Monte-Carlo time series
datasets were generated to model the gamma-ray data. Using a bin width
of one second, raw trigger rate distributions (number of triggers per
second) were compiled for each VERITAS observation run. These
distributions were used as probability density functions from which
random event times were drawn and sorted, from earliest to latest,
producing random time series with the same temporal characteristics as
the real VERITAS data. Further, to enable the measurement of the level
of any VHE flux increase seen in the VERITAS data, or to facilitate
the calculation of flux upper limits, these Monte-Carlo time series
datasets were injected with a simulated signal from the Crab pulsar as
explained in the following section.

\begin{figure*}
\centering 
\subfigure[Event Arrival Times]
{
\includegraphics[width=0.67\textwidth]{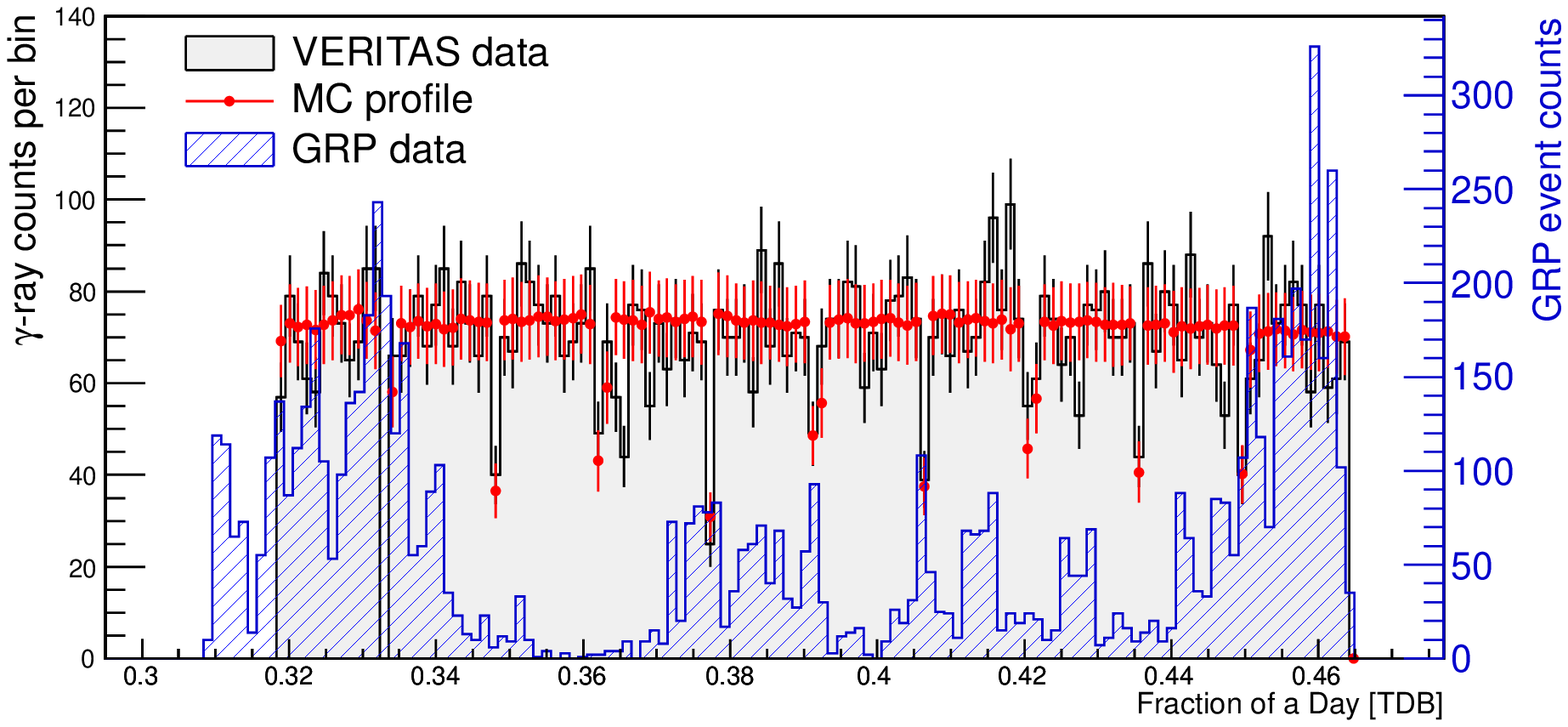}
}
\subfigure[Event Separation]
{
\includegraphics[width=0.28\textwidth]{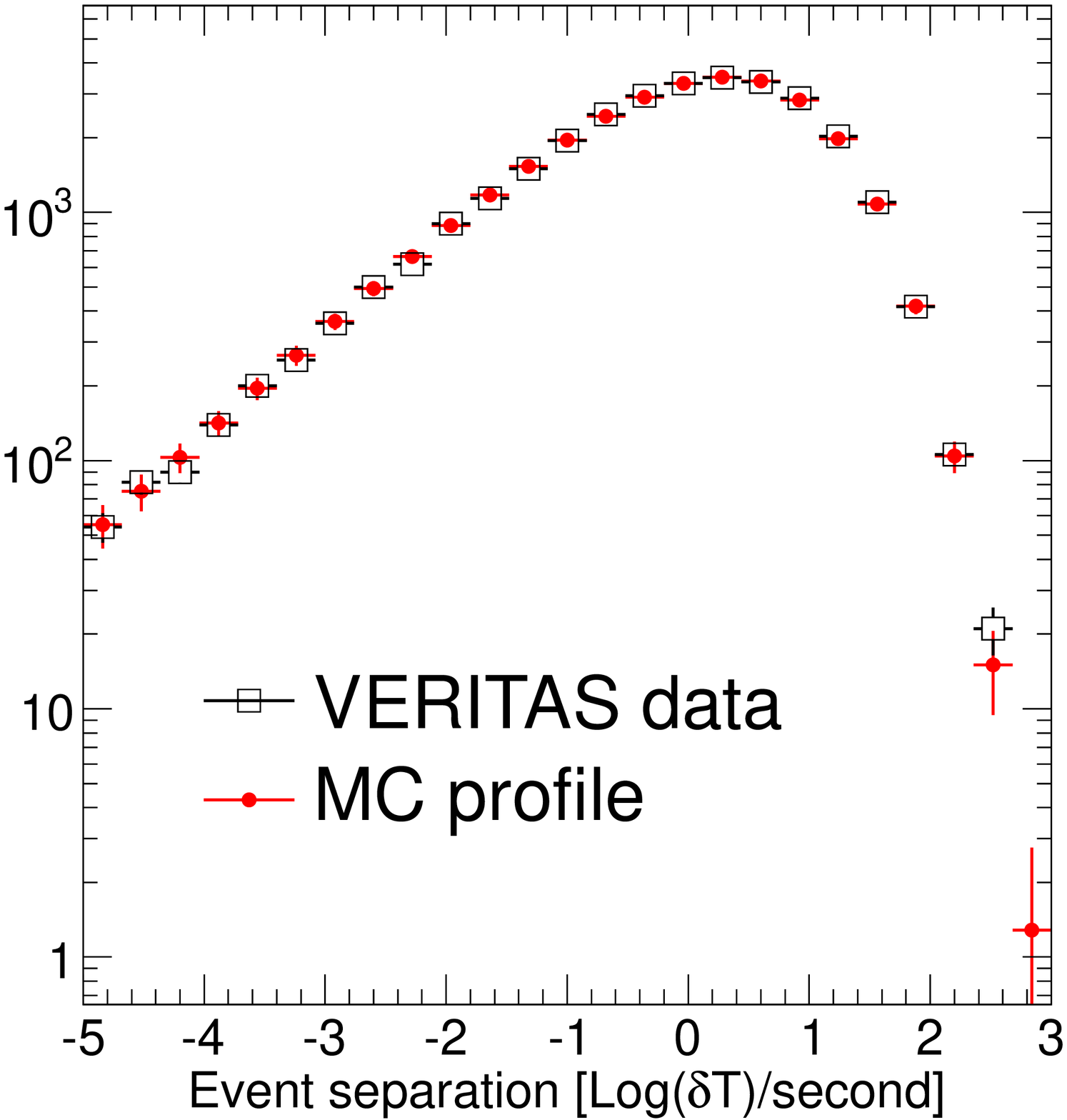}
}
\caption{Panel (a) shows the distribution of arrival times at the
  solar system barycenter of the gamma-ray (grey), radio (hatching)
  and Monte-Carlo data (circles) for the observations made on MJD
  55153. The sharp drops in the gamma-ray data rate are due to gaps in
  observation between successive 20 minute exposures. Panel (b) shows
  the arrival time event-separation distribution for each pair of
  consecutive events in the complete gamma-ray and Monte-Carlo
  datasets. In both panel (a) and (b), the Monte-Carlo data contain an
  injected signal from the Crab pulsar at the level of the measured
  VHE pulsar flux ($x_{i} = 1$) and are plotted as a profile
  histogram, with the height and error bar of each bin representing
  the mean and standard deviation determined from 500 simulated time
  series datasets respectively.}
\label{fig:MCVData1}
\end{figure*}

\begin{figure}
\centering \includegraphics[width=0.45\textwidth]{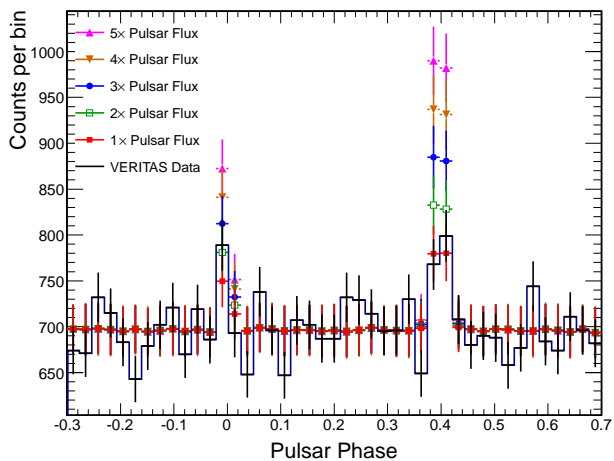}
\caption{The VERITAS Crab pulsar phasogram determined from the 11.6
  hours of simultaneous exposure together with phasograms
  determined from Monte-Carlo time series datasets. The Monte-Carlo
  data contain a simulated signal from the Crab pulsar at various flux
  levels and are plotted as a profile histogram, with the height and
  error bar of each bin representing the mean and standard deviation
  determined from 500 simulated time series datasets respectively.}
\label{fig:MCVData2}
\end{figure}

\begin{table*}[t]
\begin{center}
\begin{tabular}{cccc}
Date [MJD] & Total Number  & Number of Excess & Number of Background \\
           & $\bar{N}_{T}$ & $\bar{N}_{ex}$ & $\bar{N}_{bg}$ \\\hline
54829      & 5566          & 26.2            &  5539.8\\
55153      & 8738          & 41.1            &  8696.9\\
55158      & 7928          & 37.3            &  7890.7\\
55180      & 7861          & 37.0            &  7824.0\\\hline
Total      &30093         & 141.6            & 29951.4
\end{tabular}
\end{center}
\caption{The number of signal and background events which are present
  in the VERITAS data recorded on the four nights of observation as
  determined by the linear relationship discussed in the text. The
  number of background events is defined as the total number of events
  minus the estimated excess. When generating a Monte-Carlo time
  series which models the observations made on a given date, the
  number of background events to be generated is drawn from a Poisson
  distribution with a mean $\bar{N}_{bg}$. The number of pulsar events
  to be generated is drawn from a Poisson distribution with a mean
  $\bar{N}_{ex} \times x_{i}$, where $x_{i}$ is used to scale the
  number of excess events to the desired pulsar flux level.}
\label{tab:Inject}
\end{table*}

From our earlier analysis of the complete 107 hour VERITAS Crab pulsar
dataset, we determined that the excess of events, which fall within
the VHE main and interpulse emission regions, grows linearly with
respect to the total number of events selected
\citep{Aliu2011Sci}. This linear growth is plotted in
Figure~\ref{fig:excess}. With a total excess of 1256$\pm$130 in 267088
events we determine that there are 4.7$\pm$0.48 excess events in the
VHE emission phases per 1000 events selected. Given this linear
relationship, the number of excess events which lie within the VHE
emission phases, for any subset of the VERITAS data, can be estimated
(see Table~\ref{tab:Inject}). Further, we know that the emission peaks
in the VERITAS Crab pulsar phasogram can be modelled by two Gaussians
sitting on a uniform background. This was determined by an unbinned
maximum likelihood fit of the VERITAS phase data
\citep{Aliu2011Sci}. Using these two observations we can generate a
Monte-Carlo time series inhabited by a simulated Crab pulsar signal at
any desired flux level.

If a given VERITAS observation has a total number of events,
$\bar{N}_{T}$, using the linear relationship discussed above, the
number of excess pulsar events expected within this sample,
$\bar{N}_{ex}$, can be estimated. The number of background events is
then $\bar{N}_{bg} = \bar{N}_{T} - \bar{N}_{ex}$. Now assume we want
to model this VERITAS observation with a simulated signal from the
Crab pulsar at a flux level $x_{i}$, where $x_{i}$ is in units of the
average pulsar flux level measured with VERITAS. We draw a number of
background events, ${N}_{bg}$, from a Poisson distribution with a mean
$\bar{N}_{bg}$ and we draw a number of pulsar events, $N_{ex}$, from a
Poisson distribution with a mean $\bar{N}_{ex}\times x_{i}$. Now,
using the VERITAS raw data rate distributions, discussed earlier, we
draw $N_{T} = N_{bg} + N_{ex}$ random arrival times. A fraction of
these events, $N_{ex}/N_{T}$, is randomly selected to contain the
injected pulsar signal. This is done by shifting the arrival time of
these events to the nearest time which, when barycentered and
phase-folded, would correspond to a random phase value drawn from the
double-Gaussian function which parametrises the VERITAS Crab pulsar
phasogram. This small shift in time ($<$17 ms) has no effect on the
overall data rate characteristics of the Monte-Carlo datasets, and is
only applied to a small fraction of randomly selected events within
each dataset. This procedure was used to generate sets of Monte-Carlo
data, containing a signal from the Crab pulsar at a chosen flux level,
which model all the data rate characteristics of the real VERITAS
data, accounting for the Poisson fluctuations inherent in the VERITAS
measurement of the Crab pulsar flux. Examples of the match between the
Monte-Carlo time-series datasets and the VERITAS gamma-ray data are
shown in Figures~\ref{fig:MCVData1} and \ref{fig:MCVData2}.

\section{Results}
\begin{figure}
\centering 
\includegraphics[width=0.45\textwidth]{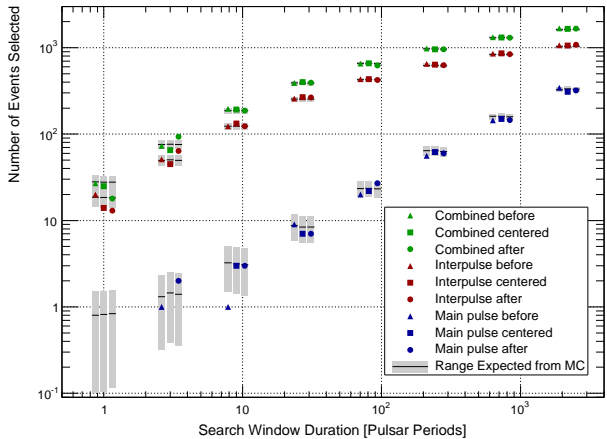}
\caption{The number of gamma-ray events selected (symbols) versus the
  duration of the search window. The grey regions denote 68\%
  containment intervals about the mean of the distribution (black line)
  determined from searches performed on the Monte-Carlo data. The
  Monte-Carlo dataset is composed of 500 simulated time series
  containing an injected signal from the Crab pulsar at the level of
  the measured VHE pulsar flux ($x_{i} = 1$). The absence of a symbol
  in 4 of the searches indicates that the number of selected events in
  each of these cases was zero. All searches return values which are
  consistent with what is found from searches on the Monte-Carlo
  datasets. The slight negative and positive shift of the x-position
  of the \textit{before} and \textit{after} symbols is done as a
  visual aid to prevent clutter about the common x-coordinate value
  for the \textit{before}, \textit{centered} and \textit{after}
  symbols.}
\label{fig:results0}
\end{figure}

\begin{figure}
\centering 
\includegraphics[width=0.45\textwidth]{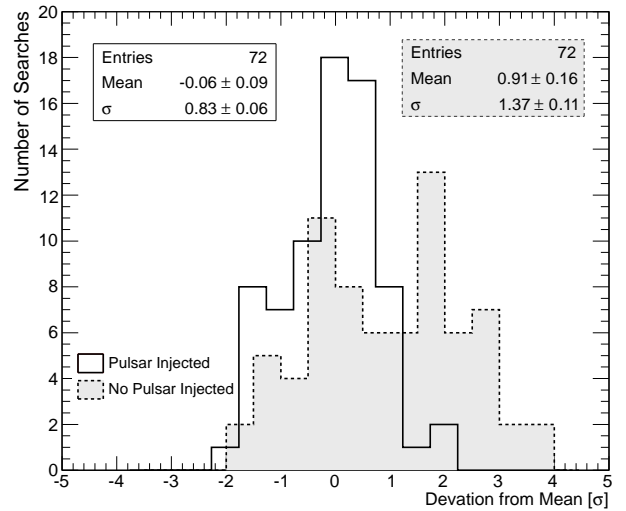}
\caption{The deviation of number of events selected in the enhancement
  searches on the real data compared to the mean number expected from
  the Monte-Carlo distributions in units of their standard
  deviation. The positive shift of the mean of the distribution
  determined from searches on Monte-Carlo datasets with no pulsar
  signal flux signal injected (grey) is due to presence of the pulsar
  signal in the VERITAS dataset and the absence of a pulsar signal in
  the Monte-Carlo. Including in the Monte Carlo a Crab pulsar signal
  at the level of the average measured VHE pulsar flux yields a
  distribution (solid black), which has a mean compatible with zero
  and a standard deviation close to one. This indicates that there is
  no significant enhancement in the gamma-ray flux correlated with
  GRPs. We note that the searches are correlated because the same
  events can be selected in different searches.}
\label{fig:results1}
\end{figure}

\subsection{Search for enhanced gamma-ray emission during GRPs}
Each enhancement search yields a number, \textit{N}, which is the
number of VERITAS events that meet the specific search criteria, i.e.,
they fall within a time interval determined by a GRP with a phase
value within the required VHE-emission phase range. The Monte-Carlo
time-series datasets were subject to the same search which was
performed on the real VERITAS dataset. Given that we generated a large
number of Monte-Carlo time series, a search on a Monte-Carlo dataset
will yield a distribution of the number of selected events, which is
approximately Gaussian, and will have a mean, $\mu$, and a variance,
$\sigma^{2} = \mu$. The number of events selected in a given search,
\textit{N}, can be compared to the mean number expected from the Monte
Carlo when no enhancement is present, $\mu$. Such a comparison is
plotted in Figure~\ref{fig:results0}, showing the number of events
selected in searches of the VERITAS data, compared to the mean number
selected from a Monte-Carlo dataset. Here, the Monte-Carlo dataset is
composed of 500 simulated time series, each containing an injected
signal from the Crab pulsar at the level of the average measured VHE
pulsar flux ($x_{i} = 1$). Using the formula, $S = (N-\mu)/\sigma$,
one can determine the statistical significance of any deviation of the
measured number of events, from the number expected in the absence of
any enhancement\footnote{Twelve of the searches on the Monte-Carlo
  datasets yield distributions which are more Poissonian in shape than
  Gaussian ($\mu \leq 10$), meaning that the strict equivalence of the
  significance formula breaks down. The resulting assertions are,
  however, unchanged.}. The significance distribution derived from the
72 searches is plotted in Figure~\ref{fig:results1}. This distribution
has a mean compatible with zero and a standard deviation close to one,
indicating the absence of any significant enhancement in the VHE
gamma-ray flux within the specified search windows positioned around
GRPs observed at 8.9 GHz.

\subsection{Calculation of the upper limit on the flux increase}
\begin{figure*}
  \centering
  \includegraphics[width=0.95\textwidth]{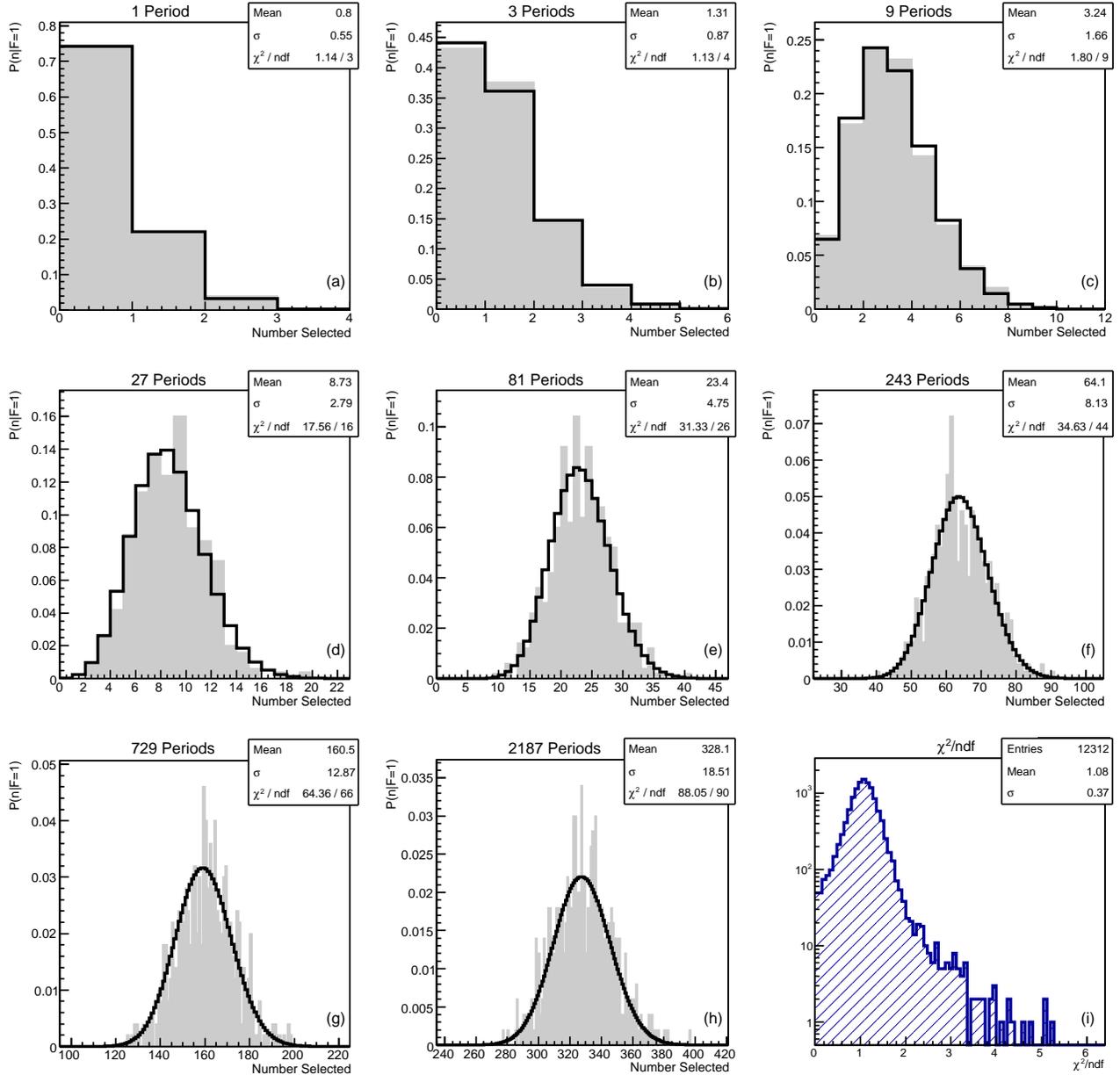}
\caption{Panels (a) to (h) show typical examples of the distribution
  of the number of selected events found in searches performed on main
  pulse GRPs. These distributions (grey) were determined from
  Monte-Carlo datasets containing an injected signal from the Crab
  pulsar at the level of the measured VHE pulsar flux.  The solid
  black curve shows the best fit Poisson curve for each
  distribution. From the fits to distributions such as these, the
  values $P(N|F=x_{i})$ can be determined. Panel (i) shows the
  distribution of reduced-$\chi^{2}$ values ($\chi^{2}/$ndf) for all
  of the Poisson fits to the Monte-Carlo distributions performed in
  this work, indicating that the Poissonian form accurately describes
  the data.}
\label{fig:fits}
\end{figure*}

\begin{figure*}
\centering \includegraphics[width=0.85\textwidth]{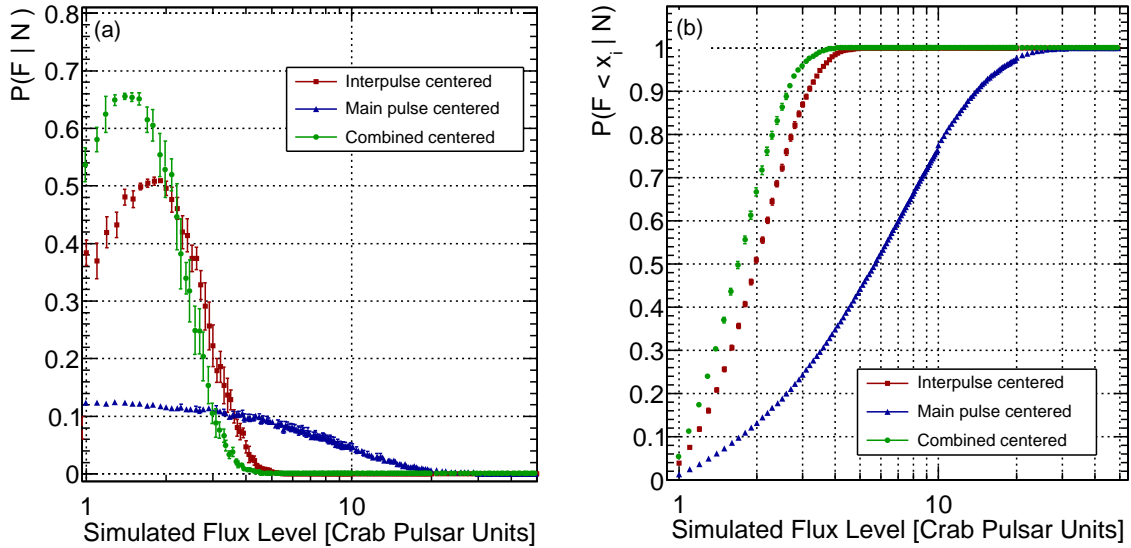}
\caption{ Panel (a) and (b) show the results of correlation searches
  centered on GRPs with search windows lasting 27 pulsar
  rotations. Panel (a) plots the posterior probability $P(F=x_{i}|N)$
  for a range of flux levels, given the number of events selected in
  the enhancement search. These probability distributions peak close
  to a flux value of 1, meaning that the flux is most likely unchanged
  during the GRP events. The error bars represent the uncertainty due
  to finite statistics in the Monte-Carlo distributions and the
  fitting procedure used to determine the probability value from the
  these distributions. Panel (b) shows $P(F < x_{i} | N )$, the
  cumulative distributions of the corresponding probability
  distributions shown in panel (a). From such curves, an upper limit
  on the level of the enhanced flux can be read at a desired
  confidence level. In the given example, the 95\% confidence level
  upper limits on the enhanced flux are 17.2, 3.5 and 2.9 times the
  flux measured by VERITAS for the main pulse, interpulse and combined
  searches respectively.}
\label{fig:pdf}
\end{figure*}

We follow the same prescription as \cite{Bilous2011} to compute the
upper limit on the VHE flux during the enhancement searches. From
Bayes' theorem, the posterior probability that the pulsar flux is $F$
given that we observed $N$ events in an enhancement search is
\begin{equation}
P(F | N ) = \frac{P(N|F)\,P(F)}
{\int\limits_{0}^{\infty} P(N|F')\,P(F')\,dF'}
\end{equation}
where $P(N|F)$ is the likelihood of selecting $N$ events in a search
when the pulsar flux is $F$. For simplicity, all flux values specified
here are cast in units of the average flux of the Crab pulsar measured
with VERITAS. $P(F)$, the prior distribution of the flux, is an
uninformative prior which we set to be
\begin{equation}
P(F) = \left\{
  \begin{array}{l l}
    C & \quad \mbox{if} \,\,\,\,1 <= F <= 50\\ 0 &
    \quad \mbox{if} \,\,\,\,F < 1, F > 50 \\
  \end{array} \right.
\end{equation}
where $C$ is a non-zero constant. This means that during the emission
of a GRP we consider the pulsar flux to be, with uniform probability,
between 1 and 50 times the average Crab pulsar flux, and to have zero
probability otherwise.

From the searches performed on the Monte-Carlo datasets we can explore
the likelihood value, $P(N|F)$, for a range of flux values, by
generating a likelihood curve. Each point in the curve is computed by
probing the distribution of the number of selected events determined
from a Monte-Carlo dataset generated with a given flux level, $F =
x_{i}$ (see Figure~\ref{fig:fits} for some example distributions). The
y-value of each point in the curve is the fraction of the Monte-Carlo
datasets which, when subjected to the specific enhancement search,
yielded the same number of coincident events as were found in the real
data. In practice, we fit the distribution yielded from searches on
the Monte-Carlo dataset with a Poisson function, and determine the
likelihood from the fitted function. This is done to minimise
fluctuations caused by the finite statistics used to compile the
Monte-Carlo distributions. Figure~\ref{fig:fits} shows some examples
of the fits to the Monte-Carlo distributions along with the
distribution of the reduced-$\chi^{2}$ values ($\chi^{2}/$ndf) for all
of the fits performed. From the reduced-$\chi^{2}$ distribution, which
has a mean value close to 1, it is clear that the Poisson functional
form accurately describes the distributions derived from searches
performed on the Monte-Carlo datasets.

The likelihood curve for each search was compiled using Monte-Carlo
datasets with injected flux levels ranging from, $x_{i}$ = 1 to 50,
with step sizes of 0.1 between 1 and 10, 0.25 between 10 and 20, and 1
between 20 and 50. Given our choice of the prior $P(F)$, and that we
evaluated the likelihood at an array of discrete flux levels,
Equation~1 for the posterior probability can be re-written as
\begin{equation}
P(F = x_{i}| N )\,\,\,=\,\,\, \frac{P(N|F = x_{i})}
{\sum\limits_{i=1}^{160} P(N|F = x_{i})\,\Delta x_{i}}
\end{equation}
where $i$ runs over the array of simulated flux levels and $\Delta
x_{i}$ is step size between each consecutive flux level. This
posterior probability curve is normalised and can be integrated
yielding a cumulative posterior probability, $P(F < x_{ul}| N )$, from
which one can determine the flux upper limit, $x_{ul}$, at a chosen
confidence level. Figure~\ref{fig:pdf} shows the posterior probability
and cumulative distribution functions for three different
searches. The flux value where the cumulative probability distribution
crosses 0.95, marks the upper limit on the emitted flux correlated
with GRPs at the 95\% confidence level.

\begin{figure}
\centering \includegraphics[width=0.45\textwidth]{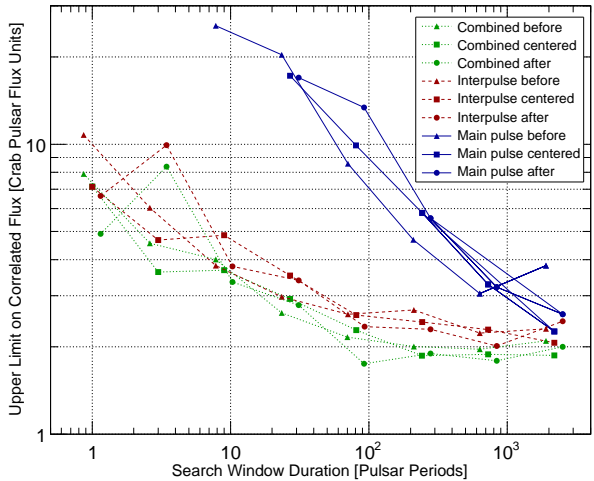}
\caption{95\% confidence level upper limit on the gamma-ray flux
  correlated with GRPs for 64 of the 72 searches. We do not quote an
  upper-limit value for 8 searches with short windows placed around
  main pulse GRPs, but note that it is likely around or above 50 Crab
  pulsar flux units. The slight negative and positive shift of the
  x-position of the \textit{before} and \textit{after} symbols is done
  as a visual aid to prevent clutter about the common x-coordinate
  value for the \textit{before}, \textit{centered} and \textit{after}
  symbols.}
\label{fig:UL}
\end{figure}

The upper limits on the flux correlated with GRPs at the 95\%
confidence level are plotted in Figure~\ref{fig:UL} for 64 of the 72
searches. On the shortest time scales probed, the duration of one
pulsar period, a limit of $\sim$5-10 times the average Crab pulsar
flux is set on the inter-pulse and the combined inter-pulse and main
pulse searches. For 8 of the searches performed around main pulse
GRPs, the posterior probability curve had not converged to zero before
a flux value of 50 times the measured flux. This is due to the low
rate of main pulse GRPs detected at 8.9 GHz, which resulted in a small
number of selected gamma-ray events in some of the main pulse GRP
searches. The relatively large fluctuations inherent in small-number
statistics result in wide posterior probability curves and, thus, very
high upper limit values.  We do not quote a 95\% confidence level
upper-limit value for these searches, but note that it is likely
around 50 times the average gamma-ray flux. As the search window size
is increased, and thus the statistical sample increases, these
fluctuations decrease and the upper-limit values become more
constraining.  Once enough of the gamma-ray sample is selected due to
the increasing size of the search windows, the upper-limit values
level out at $\sim$2-2.5 times the average Crab pulsar flux. We note
that for the longest two search windows, 729 and 2187 periods, between
70\% and 80\% of the VERITAS events are selected in the interpulse and
combined searches resulting in a reduced enhancement sensitivity in
these searches \footnote{The 70\% and 80\% values were determined
  without inclusion of the data acquired on the first day of
  observations where our GRP count was relatively low.}.

\subsection{Estimation of Uncertainty}
As stated earlier, the fraction of Monte-Carlo searches which yield
the same number of correlated events as were found in the data,
$P(N|F)$, was determined from a Poisson fit to the distributions,
rather than from the distributions themselves. By adopting this
procedure we found that fluctuations in the posterior probability
density curves are dramatically reduced, while having little effect on
the computed 95\% confidence level upper-limit values. We investigated
the uncertainty on the posterior probability values due to the finite
statistics in the Monte-Carlo distributions and the Poisson fitting
procedure.

The $\chi^{2}$ values were determined between a given Monte-Carlo
distribution and Poisson curves with a range of mean values centered
on the best fit Poisson mean value. From these $\chi^{2}$ values, a
fit probability versus Poisson mean curve was generated, and from this
curve, random values were drawn. A histogram of $P(N|F)$ values was
then compiled using Poisson curves with these random values as their
mean. The standard deviation of this histogram was then used as the
uncertainty on the $P(N|F)$ value determined from the best fit Poisson
curve. The uncertainty on the probability values determined in this
way is less than 15\% for the bulk of distributions but is
occasionally as large as 30\%. The error bars on the probability
values plotted in Figure~\ref{fig:pdf} were calculated in this
way. Other methods to calculate the uncertainty were investigated and
found, in general, to yield a smaller uncertainty value. Folding the
uncertainty on the probability values into the cumulative
distribution, however, the uncertainly on the 95\% confidence level
upper-limit value was found to be less than 3\% for the 64 searches
which yielded a limiting value. This underlines the robustness of the
procedure we used to calculate the upper-limit values.

\subsection{Selecting only the most energetic GRP events}
Having investigated GRPs whose peak flux density is greater than
7$\sigma$ above the averaged radio signal and observed no VHE
enhancement, we now consider only the most energetic GRP events. Due
to the low rate of GRP events in the 2008 data set, the following
analysis uses the data collected in 2009 only.

Earlier, we estimated that the system equivalent flux density for the
2009 observations was 0.7~Jy. The sampling interval for these
observations was 3.2~$\mu$s. Thus, the energy of a GRP is
\begin{equation}\label{eqn:GRPE}
GRP_\mathrm{E} = \frac{GRP_\mathrm{S/N}}{\sqrt{N_\mathrm{samp}}}(0.7~\!\mathrm{Jy}) \times (N_\mathrm{samp})(3.2~\!{\mu}s)
\end{equation}
where $GRP_\mathrm{S/N}$ is the signal-to-noise ratio of the GRP and
$N_\mathrm{samp}$ is the width of the GRP in samples (see
\cite{Bilous2012}, for more details regarding the energy calculation
of GRPs). The effective width of each GRP is determined from the width
of the boxcar function which, when convolved with the radio time
series, returns the largest S/N value for a given pulse. The widest
GRP found had a width of 14 samples, or $44.8~\mu$s.

\begin{figure}
\centering \includegraphics[width=0.45\textwidth]{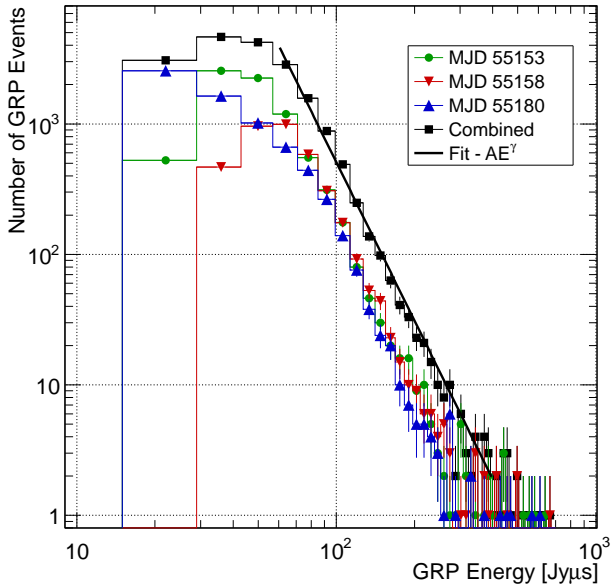}
\caption{The distribution of giant pulse energies for the latter three
  nights of observations.  The distributions follow a power law, with
  the combined data set having a spectral index of $-4.03$. The
  different number of GRPs selected below the $\sim$60~Jy$\mu$s
  roll-off on each night of observation is due to the different amount
  of RISS contributions to the GRP flux densities on a given night.}
\label{fig:Esummary}
\end{figure}

In order to have a common energy scale for observations taken on
different nights, we must correct for the effects of RISS. This was
done by compiling average radio pulse profiles on 80-minute time
scales, which is the characteristic time scale of RISS for the Crab at
8.9 GHz. The profile for the second half of the observing session on
MJD 55180 was the strongest and we chose it as reference pulse profile
to which the other profiles were compared. Ratios of the peak S/N
values were computed between the reference profile and the other
profiles and used as RISS-correction coefficients to scale the GRP
energy values computed by
Equation~\ref{eqn:GRPE}. Figure~\ref{fig:Esummary} shows the resulting
distributions of GRP energy, which follow the expected power-law shape
with a spectral index of $-4.03$.  The roll-off below
$\sim$60~Jy$\mu$s is caused by a bias in our GRP selection, which was
based on peak fluxes and reduces the number of broad weak pulses
selected. The different number of GRPs selected below the roll-off on
each night of observation is due to the different amount of RISS
contributions to the GRP flux densities on a given night.

\begin{figure*}
\centering 
\includegraphics[width=0.99\textwidth]{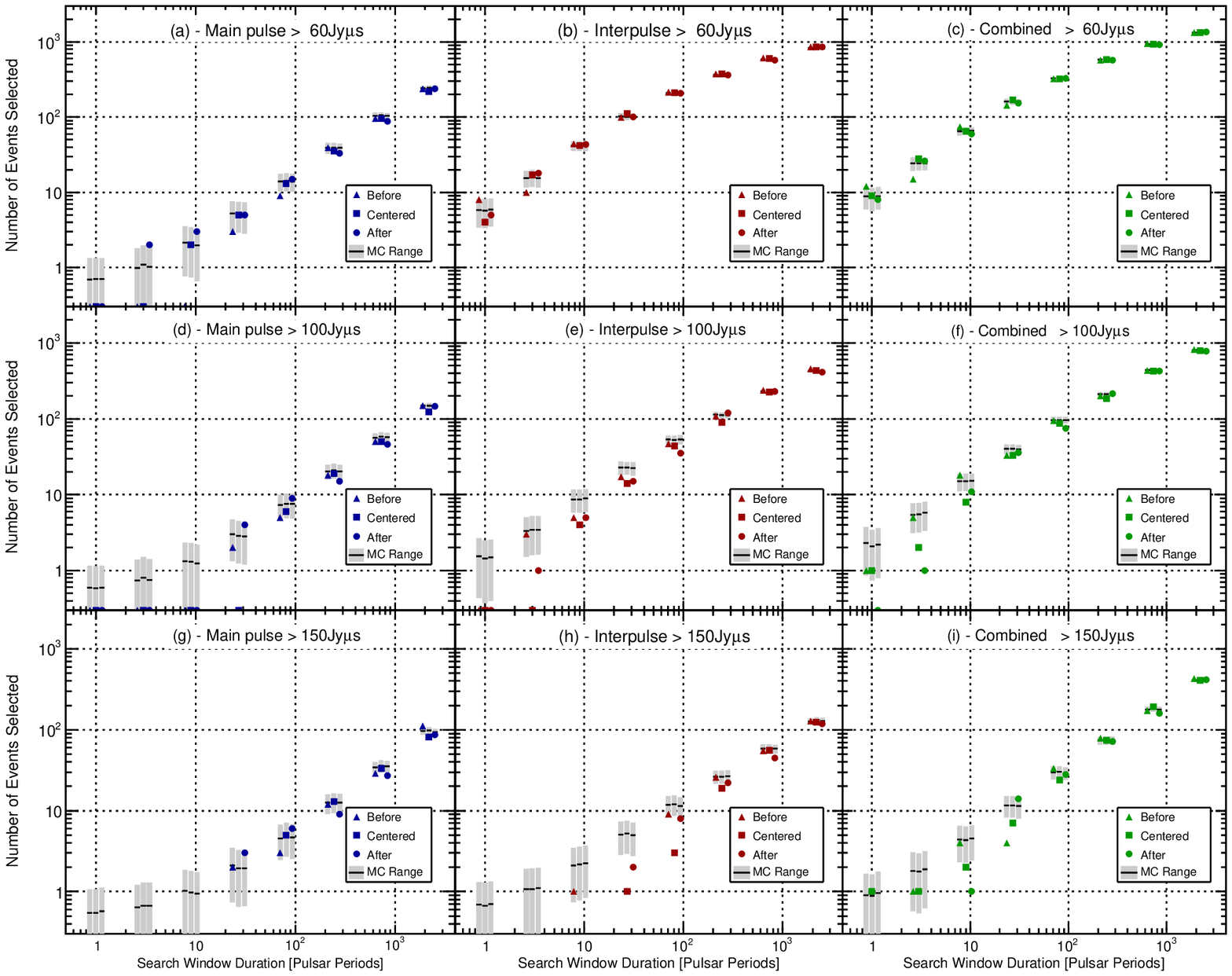}
\caption{The number of gamma-ray events selected (symbols) versus the
  duration of the search window. The grey regions denote 68\%
  containment intervals about the mean of the distribution (black line)
  determined from the identical searches performed on Monte-Carlo data
  sets containing an injected signal from the Crab pulsar at the level
  of the measured VHE pulsar flux. The upper, middle and lower rows
  show the results of the enhancement search when restricting the
  energy of the GRP events to be above 60~Jy$\mu$s, 100~Jy$\mu$s and
  150~Jy$\mu$s, respectively. No excess (or deficit) is found in any
  search with a probability equivalent of $3\sigma$ or higher. See the
  caption of Figure~\ref{fig:results0} for a further description of
  this figure.}
\label{fig:results0cut}
\end{figure*}

Having computed an RISS-corrected energy value for every GRP event in
the 2009 data set, we repeated the correlation search with the VERITAS
data. We applied three different energy cuts to the radio events,
namely: 60~Jy$\mu$s, 100~Jy$\mu$s, and 150~Jy$\mu$s. The 60~Jy$\mu$s
threshold was chosen as a low common energy threshold which provides
an unbiased sample of GRPs, independent of both the GRP search bias
and the RISS conditions during a particular observing session. The
other two thresholds were chosen to be ``high'' and ``very high'',
selecting the most energetic $\sim$6\% and $\sim$1\% of GRPs,
respectively. The results of these searches are plotted in
Figure~\ref{fig:results0cut}. We find no significant enhancement
($>3\sigma$) in the VHE emission correlated with these energetic GRP
events.

\section{Discussion and Conclusion}
Following our study of simultaneous radio and gamma-ray data we
observe no significant enhancement in VHE gamma-ray emission from the
Crab pulsar correlated with GRPs observed at 8.9~GHz. Our findings are
similar to those previously reported at lower gamma-ray and X-ray
energies \citep{Lundgren1995, Bilous2011, Bilous2012}. Given the level
of uncertainty in theories of GRP emission, it is hard to draw firm
conclusions resulting from the lack of any observed correlation with
VHE emission. We are not aware of any theory with quantitative
predictions of correlated emission between GRPs and VHE
emission. GRP-emission mechanisms associated with changes in plasma
coherence will not cause enhancements in incoherent emission. Small
and localised changes in the pair-creation rate, which can explain the
small (3\%) optical enhancements previously measured by
\cite{Shearer2003}, would yield VHE flux enhancements which are below
our sensitivity.

Enhanced gamma-ray emission in connection with high-frequency
interpulse GRPs, as postulated by \cite{Lyutikov2007}, is not
observed. In this model the primary beam, which generates coherent
GRPs through anomalous cyclotron resonance, also emits curvature gamma
rays with energies in the tens of GeV range. It has since been
realised that there is a solid upper limit on the energy of curvature
radiation in the Crab, $E_{\gamma} \leq 150$~GeV
\citep{Lyutikov2011}. This is due to the equivalence of the rate of
energy loss and rate of acceleration gains achieved by charged
particles energised in the outer magnetosphere. Thus, in the Crab
pulsar, the curvature photons from the primary beam generally do not
reach the VHE energy band. This means that the non-detection of a VHE
enhancement, presented here, does not contradict this GRP emission
model.

However, it is possible that there is an indirect link between VHE and
radio emission for interpulse GRPs.  Recent models for the pulsed VHE
gamma-rays detected from the Crab suggest that inverse-Compton
emission dominates curvature emission at energies above a few GeV in
the outer magnetosphere \citep{Lyutikov2011,Du2012}. A recent study of
the Geminga pulsar also supports this scenario
\citep{Lyutikov2012G}. In these models, the primary particle beam is
accelerated in a modest electric field (of few percent of the magnetic
field strength) in the outer magnetosphere and produces curvature
emission up to $\sim$10 GeV. Photons in the VHE band are generated by
secondary pairs which up-scatter their own cyclotron or synchrotron
emission.  Within the framework of the high-frequency interpulse GRP
emission model of \cite{Lyutikov2007}, a VHE-GRP connection is thus
still expected. In this model, roughly half of the energy of the
plasma beam, energised in the magnetic reconnection event, goes into
the production of a secondary plasma which can in turn generate
enhanced VHE gamma-ray emission via synchrotron self-Compton
scattering. The level of the VHE enhancement is, however, difficult to
estimate given the uncertainties and fluctuations in the properties of
both the primary and secondary beams.  This enhancement is likely
below the sensitivity of present VHE instruments.

In the model of pulsed VHE emission from the Crab pulsar of
\cite{Aharonian2012}, where pulsed X-ray emission originating in the
magnetosphere is up-scattered to VHE energies in the wind beyond the
light cylinder, a GRP-VHE correlation should exist at most at the same
level as a GRP-X-ray correlation. Since no GRP-X-ray correlation is
seen \citep{Bilous2012} and the upper limit on the enhanced X-ray flux
is 30\%, any corresponding VHE enhancement is expected to be below
30\% and thus below our sensitivity.

\acknowledgments This research is supported by grants from the
U.S. Department of Energy Office of Science, the U.S. National Science
Foundation and the Smithsonian Institution, by NSERC in Canada, by
Science Foundation Ireland (SFI 10/RFP/AST2748) and by STFC in the
U.K. We acknowledge the excellent work of the technical support staff
at the Fred Lawrence Whipple Observatory and at the collaborating
institutions in the construction and operation of the instrument. The
National Radio Astronomy Observatory is a facility of the National
Science Foundation operated under cooperative agreement by Associated
Universities, Inc.


\end{document}